\documentstyle[12pt,epsf,epsfig]{article}

\newcommand{\tr}{\mbox{Tr}}

 \let\be=\beta

\def\0{\over } 
\def\1{\vec }     
\def\2{{1\over2}} 
\def\4{{1\over4}}            
\def\5{\bar }  
\def\6{\partial } 
\def\7#1{{#1}\llap{/}}                         
\def\8#1{{\textstyle{#1}}}
\def\9#1{{\bf {#1}}}                           
 
\def\llp{\hbox to 0pt{\hss/\hskip1.5pt}}
\def\llo{\hbox to 0.2pt{\hss /}} \def\llq{\hbox to 0pt{\hss/\hskip0.5pt}}
\def\so{\supset\hbox to 0pt{\hss $\displaystyle -$\hskip1pt}}

\let\nn=\nonumber  
\def\bea{\begin{eqnarray}} \def\eea{\end{eqnarray}} 
\def\beann{\begin{eqnarray*}} \def\eeann{\end{eqnarray*}} 
\def\beq{\begin{equation}} \def\eeq{\end{equation}}  
\begin{document} 
\setlength{\baselineskip}{18pt}                                     
\thispagestyle{empty}
\begin{flushright}
{\tt CERN-TH 99-413\\ December 1999}
\end{flushright}
\vspace{5mm}
\begin{center}
{\Large \bf
 Adjoint String Breaking \\ in 4d SU(2) Yang-Mills Theory }\\
 \vspace{15mm}
\renewcommand{\thefootnote}{\fnsymbol{footnote}}
{\large Philippe de Forcrand$^{1,2}$ 
 and Owe Philipsen$^{2}$}\\
\renewcommand{\thefootnote}{\arabic{footnote}}
 \vspace{10mm}
{\it 
$^{1}$ ETH-H\"onggerberg, CH-8093 Z\"urich, Switzerland\\
$^{2}$ CERN-TH, CH-1211 Geneva 23, Switzerland}

\end{center}
\vspace{2cm}

\begin{abstract}
\thispagestyle{empty}
\noindent
We compute the static potential of adjoint sources in SU(2) Yang-Mills
theory in four dimensions by numerical Monte Carlo simulations. 
Following a recent calculation in 2+1 dimensions,
we employ a variational approach involving string and gluelump operators and
obtain clear evidence for string breaking and the saturation of the potential
at large distances. For the string breaking scale we find 
$r_b\approx 1.25{\rm fm}, 2.3 r_0$, or in units of the lightest glueball,
$r_b m_{0++}\approx 9.7$.
We furthermore resolve the first 
excitation of the flux-tube and observe its breaking as well. 
The result for $r_b$ is in remarkable quantitative agreement  
with the three-dimensional one.
\vspace*{3cm}
\flushleft{CERN-TH 99-413}
\end{abstract} 
\setcounter{page}{0}

\newpage

\section{Introduction}

The most essential property of QCD is confinement: the energy of a static
quark-antiquark $Q\bar{Q}$ pair grows linearly with their separation $r$,
because the colour flux between the two charges is negligible outside 
a thin tube, the string, whose energy grows in proportion to its length.
This remains true up to a distance $r_b$ where the string breaks: the energy 
becomes sufficient to create a pair of mesons, $Q\bar{q}$ and 
$\bar{Q}q$, which can be taken arbitrarily far apart at no extra energy
cost. The static potential thus levels off at a value $2 E(Q\bar{q})$.
Similarly, the force between a pair of adjoint sources $Q_{adj}$ is screened 
by gluons $g$, so that the adjoint string breaks, even in the absence of quarks,
when a pair of ``gluelumps''
$Q_{adj}g$ can be created. 
In numerical simulations of lattice QCD
it has been far easier to observe the string itself
than its breaking. In spite of extensive searches for colour screening
\cite{QCD_string}, 
the evidence is still preliminary at best \cite{break}.

The essential reason lies in the projection properties 
of the Wilson loop, whose change from an
area law to a perimeter law behaviour would signal string breaking.
The difficulty with seeing the perimeter law
in simulations stems from the fact that
at $r>r_b$ the state with unbroken string between the sources is
still an eigenstate of the Hamiltonian, albeit an excited one.
By construction, the Wilson loop has good overlap with this state,
which remains the case also for $r>r_b$. If it also
has poor overlap with the two-gluelump state, then for $r>r_b$
the latter, which is the true ground state, will only dominate
the correlation function for very large correlation times $t$,
which according to the definition of the potential 
should be taken to infinity for the loop
to yield the ground state.
In most practical simulations, the limit $t\rightarrow
\infty$ is not realized, but $t$ is typically a few lattice
spacings when the signal is lost in noise. If this happens before
$t$ is large enough for the ground state to dominate the
correlation function, the ground state will be missed and an excited state
will be extracted instead, leading to the apparent
conclusion that the string does not break \cite{unbroken}.

A way to circumvent this difficulty is to supplement the calculation
with an additional operator having good projection onto the
two-gluelump state. This enables an unambiguous identification
of the ground state also for $r>r_b$, even if the perimeter
behaviour of the Wilson loop is hidden in noise.
After a first attempt for the adjoint potential \cite{mic92},
this procedure has been first sucessful for the fundamental
static potential of the SU(2) Higgs model in its confining phase
in (2+1) \cite{sb1} as well as in (3+1) \cite{sb2} dimensions,
where it clearly demonstrated the flattening of the potential
as well as mixing between the Wilson loop and the 
broken string state in a very narrow region around $r_b$.
The same findings were reported from simulations of the adjoint
potential in (2+1) dimensional Yang-Mills theory \cite{ps,pw1}.

In this work, we study string breaking in the adjoint potential of
SU(2) Yang-Mills theory in four dimensions.
In practical terms, the advantage over QCD is that 
no dynamical quarks need be considered. The price to pay is that the adjoint
string tension is larger than the fundamental one, approximately by the
ratio 8/3 of the Casimirs, so that Wilson loops obeying 
an area law disappear more quickly in the statistical noise, which 
turned out to hamper the earlier study \cite{mic92}. 

Section 2 describes our calculational method for on- and off-axis correlations.
Section 3 presents our simulations and Section 4 our numerical results. 
Section 5 contains a summary and outlook.

\section{Calculational method}

We work with the Wilson action
\beq
S[U]= \beta_G \sum_x \sum_{\mu<\nu}\left[1-\frac{1}{2} \tr P_{\mu\nu}(x) \right],
\eeq
where $P_{\mu\nu}$ denotes a plaquette of links $U_\mu$ in the fundamental
representation, $\be_G=4/g^2$ and $g$ is the bare gauge coupling.
The links in the adjoint representation, $A_\mu^{ab}$,
are related to the fundamental $U_\mu$'s by
\beq
  A_\mu^{ab}(x) = \frac{1}{2}\,{\tr\,}\left(\sigma^a U_\mu(x)\sigma^b
  U^\dag_\mu(x)\right),
\eeq
where $\sigma^a$ are the Pauli matrices. 

The standard operators to extract the static potential are rectangular
Wilson loops with side lengths $r$ and $t$ in a spatial and the time
direction, respectively. Physically, a Wilson line $S(x,y)$ is interpreted
as colour flux propagating along that line from $x$ to $y$.
Accordingly, a straight Wilson line in $t$-direction corresponds to
a static colour source, whereas a Wilson line in $r$-direction may be 
used to describe a flux tube between $x$ and $y$.
A rectangular Wilson loop then represents the correlation function
of a string of length $r$ over a time interval $t$. 
To consider the case of adjoint sources, 
an adjoint Wilson loop is needed, which 
takes the form
\beq
\label{wl}
G_{SS}(r,t)=W_{\rm adj}(r,t)= \mid W(r,t) \mid^2 -1,
\eeq
where 
\beq
\label{wl2}
W(r,t)=\tr\left[S(0,r\hat{\j})S(r\hat{\j},r\hat{\j}+t\hat{0})
S^{\dag}(t\hat{0},r\hat{\j}+t\hat{0})S^{\dag}(0,t\hat{0})\right]
\eeq
is the fundamental Wilson loop, $S(x,y)$ is a 
straight line of fundamental links connecting the sites $x$ and
$y$, and $\hat{j}$ and $\hat{0}$ denote unit vectors in a spatial and 
the time direction respectively.  
The static potential in the adjoint representation is 
defined in terms of the exponential decay of the adjoint Wilson loop,
\beq \label{defpot}
V(r)=-\lim_{t\to\infty}\frac{1}{t} \ln[W_{\rm adj}(r,t)].
\eeq
In the region of linear confinement the Wilson loop obeys an area
law, whereas for distances beyond some breaking scale $r_b$ a perimeter
law is expected. This perimeter behaviour for 
large separations $r$, however, has not been observed to date.

The correlation function for a bound state of a static adjoint colour
source and a gluon field, called gluelump in the literature
\cite{mic85}, is given by the non-local gauge-invariant operator
\bea \label{lump}
   G_G(t) &=& \left\langle \tr ( P(x) \sigma^a) S_{adj}^{ab}(x,x+t\hat{0})
               \tr (P^{\dag}(x+t\hat{0})\sigma^b)\right\rangle \nn \\ 
          &=& \left\langle \tr \left[ P(x)S(x,x+t\hat{0}) \left( P^{\dag}(x+t\hat{0})-
                P(x+t\hat{0}) \right) S^{\dag}(x,x+t\hat{0})\right]\right\rangle
\eea
with the adjoint representation Wilson line
\beq
S_{adj}^{ab}(x,y)=\frac{1}{2}\tr \left (\sigma^a S(x,y) \sigma^b
  S^{\dag}(x,y) \right) .
\eeq
Here $P(x)$ denotes the ``clover-leaf'' of all four
plaquettes with the same orientation which emanate from the
endpoint~$x$ of the adjoint Wilson line, summed over the three spatial
planes ~\cite{mic92}. 

We now follow the procedure proposed in \cite{mic92} and 
applied in three dimensions \cite{pw1} to construct an
operator projecting on two of these bound states at distance $r$ by
\bea \label{mm}
  G_{GG}(r,t) &=&
  \left\langle \tr \left[ P(0)S(0,t\hat{0}) \left( P^{\dag}(t\hat{0})-
  P(t\hat{0}) \right) S^{\dag}(0,t\hat{0})\right] \right. \\
  & &\times \left .
  \tr \left[ P(r\hat{\j})S(r\hat{\j},r\hat{\j}+t\hat{0}) 
      \left( P^{\dag}(r\hat{\j}+t\hat{0})-P(r\hat{\j}+t\hat{0})
      \right) S^{\dag}(r\hat{\j},r\hat{\j}+t\hat{0}) 
         \right]\right\rangle . \nn
\eea
Finally, correlations between a string and a gluelump state,
and vice versa, may be described by
\bea
 G_{SG}(r,t) &=& \Big\langle{\rm Tr\,}\Big [
 \left(P^{\dag}(t\hat{0})-P(t\hat{0})\right)\,S^{\dag}(0,t\hat{0})\,
 S(0,r\hat{\j}) S(r\hat{\j},r\hat{\j}+t\hat{0})\nn \\
 & & \times P(r\hat{\j}+t\hat{0})
 S^{\dag}(r\hat{\j},r\hat{\j}+t\hat{0})
S^{\dag}(0,r\hat{\j})S(0,t\hat{0})\Big ]
 \Big\rangle ,\nn\\ 
 G_{GS}(r,t) &=& \Big\langle{\rm Tr\,}\Big [
 \left(P^{\dag}(0)-P(0)\right)\,S(0,t\hat{0})\,
 S(t\hat{0},r\hat{\j}+t\hat{0}) S^{\dag}(r\hat{\j},r\hat{\j}+t\hat{0})\nn \\
 & & \times P(r\hat{\j})
 S(r\hat{\j},r\hat{\j}+t\hat{0})
S^{\dag}(t\hat{0},r\hat{\j}+t\hat{0})S^{\dag}(0,t\hat{0})\Big ]
 \Big\rangle  \;.
\label{mix}
\eea
The static potential, its excitations and the mixing between gauge string 
and two-gluelump state can then be extracted from measurements of the 
matrix correlator
\beq
 G(r,t) = \left(\begin{array}{ll}
        G_{SS}(r,t) & G_{SG}(r,t) \\
        G_{GS}(r,t) & G_{GG}(r,t)
                \end{array}\right).
\label{eq_matcor}
\eeq
We also keep the single gluelump operator $G_G$ from (\ref{lump}) in
our simulations, in order to check whether $G_{GG}$ defined in
(\ref{mm}) indeed has a good projection onto a two-gluelump state, for
which one expects $E_{GG}\simeq2E_G$. 

A particular difficulty in observing string breaking comes from its suddenness.
As the distance between the sources increases, the ground state switches
abruptly from an unbroken string overlapping almost exclusively with the
Wilson loop, to a broken string overlapping almost exclusively with two
gluelumps. Mixing among the two states, which proves that one passes
continuously from one ground state to the other, occurs only if the two
states are very close in mass. This corresponds to a very narrow window of
distances, which becomes elusive or entirely unobservable on a coarse lattice
\cite{mic92}. For this reason, we supplement our study of on-axis
correlations as above with that of off-axis correlations, which offer far
more discrete Euclidean distances. 
To avoid computing spatial parallel transporters
from $x$ to $y$ as in eqs.(\ref{wl2},\ref{mm},\ref{mix}) for off-axis 
separations, we instead fix
the gauge and measure in a fixed gauge the generic 
adjoint correlator between two time-like ``segments''
\beq
G_{fix}(r,t)=\langle \,\mid \rm{Tr} S(0,t\hat{0}) 
S^\dagger(r\hat{j},r\hat{j}+t\hat{0}) \mid^2 - 1 \rangle \,.
\eeq
Any gauge which preserves the transfer matrix is acceptable. The choice
of gauge will determine the
projection of the above correlator onto the string or the two-gluelump state.
To project onto the broken string, $G_{fix}\sim G_{GG}$,
we simply choose the gauge (defined up to an irrelevant Abelian rotation
$\rm{exp}(i \theta \sigma_3)$) which brings
$F = \frac{1}{2 i} (P - P^\dagger)$
to the colour direction $\sigma_3$ at every point:
$F^1(x) = F^2(x) = 0 ~\forall x$.
The required local gauge transformation can be determined and applied
independently at each lattice site.
To project onto the unbroken string, $G_{fix} \sim G_{SS}$, 
we need a smooth gauge and choose the adjoint Coulomb gauge. 
Correspondingly, for the mixed elements $G_{SG}$ the gauges 
on the two timeslices in the correlator $G_{fix}$ are different.
In order to avoid possible systematic effects from lattice
``Gribov'' copies, we discard the usual implementation of Coulomb gauge on
the lattice by iterative local extremization \cite{Mandula} in favour of
the unambiguous prescription of Vink and Wiese \cite{VW}. Gauge-fixing is
accomplished by rotating the two lowest-lying eigenvectors of the adjoint 
covariant $3d$ Laplacian to direction $\sigma_3$ and into the plane
$(\sigma_1,\sigma_3)$ respectively in colour space.
This gauge condition corresponds to the Laplacian Center Gauge described
in \cite{Pisa}, with only the three spatial directions used for the Laplacian.
It reduces to Coulomb gauge for the adjoint field in the continuum limit.
It is unambiguous except when either of the lowest two eigenvalues 
of the Laplacian becomes
degenerate, which signals a genuine (not a lattice artifact) Gribov copy.
Such a degeneracy did not occur over the whole course of our simulations.
The computation of the lowest-lying Laplacian eigenvectors is performed with
the use of the ARPACK package \cite{arpack}. Note that it requires no more
computing effort than the usual iterative procedure.

Nevertheless, gauge-fixing remains expensive. In principle, we would like to
perform a variational study over a variety of gauges.
To keep computer resources
within our budget, we had to limit ourselves to the two gauges above, where
the links used to construct $F$ or the covariant Laplacian have been
smeared 10, respectively 20 times. The price we pay for such a truncation of
the variational basis is a poorer determination of the ground state.
Using smeared rectangular Wilson loops yields more accurate results at cheaper
cost. The specific purpose here is to exhibit the progressive mixing of
the two variational states with distance.

The energy levels extracted from the matrix correlator $G(r,t)$ in
eqn.(\ref{eq_matcor}) are linearly divergent with decreasing lattice
spacing and hence do not have a continuum limit. This divergence is
due to the self-energy of the static sources which, although 
perturbatively computable, cannot be absorbed by renormalization into
a parameter of the theory. It reflects the fact that the static
potential is defined only up to an arbitrary (infinite) constant, and
does not itself constitute a finite physical quantity. On the other
hand, the confining force and the string breaking scale 
are defined by energy differences such that the divergence cancels out,
and these quantities do have a finite and physical continuum limit. 

\section{Simulation and analysis}

In order to improve 
the projection properties of our operators, we 
employed the standard smearing algorithm \cite{alb} to obtain smeared
spatial link variables of unit length, which were then used instead of
the original ones in constructing the correlation functions defined
above. All links in the time directions were left unsmeared such that
the transfer matrix remains unaffected by our smearing procedure. As a
basis of operators used in the matrix correlator
eq.~(\ref{eq_matcor}), we chose two different link smearing levels
for the spatial Wilson lines and one smearing level with good
projection for the clover-leaves $P$. 
Thus, our correlator $G(r,t)$ represents a $3\times 3$ matrix
for the gauge-invariant calculation. In the case of the off-axis,
gauge-fixed correlators we kept only one smearing level for the
unbroken string, giving a $2\times 2$ matrix for $G(r,t)$.

The procedure we follow to diagonalize $G(r,t)$ by means of a
variational calculation has been described in detail in the literature
\cite{rak,us}, and its application to the calculation of the adjoint
potential has already been discussed \cite{mic85,mic92,ps,pw1}. Here we
use the same algorithm as in \cite{pw1} and for completeness
briefly repeat the outline
of the procedure. 
Each element $G_{ij}$ of the matrix correlator represents a correlation
function which can be written as
\beq
 G_{ij}(r,t)=\langle \phi_i(t) \phi_j(0)\rangle \;,
\eeq 
where the $\phi_i$'s represent a spatial gauge string of length $r$ at
a given smearing level for $i=1,2$, and a two-gluelump operator
at a different smearing level for $i=3$. In a temporal gauge, where
all time-like links are unity, this expression describes indeed
the measured correlation functions. 
Since all $G_{ij}$ are manifestly
gauge invariant, this remains true without gauge fixing. 
The variational diagonalization of $G(r,t)$ consists of solving the
generalized eigenvalue problem \cite{mic85,lw}
\beq
G(r,t)\,v^i(t,t_0)=\lambda_i(t,t_0)G(r,t_0)v^i(t,t_0),\quad
t>t_0.
\eeq
From the eigenvectors $v^i$ one may construct the corresponding eigenstates
\beq
  \Phi_i = c_i\sum_k v_k^i\phi_k=\sum_k a_{ik} \phi_k,
\label{eq_eigenstate}
\eeq
which are superpositions of the string and gluelump operators used in
the simulation. The constants $c_i$ are chosen such that $\Phi_i$ is
normalized to unity. The diagonalized correlation matrix may then be
written as
\beq \label{diag}
\Gamma_i(r,t) = \langle \Phi_i(t) \Phi_i(0) \rangle =
\sum_{j,k=1}^3\,a_{ij}a_{ik}\,\phi_j(t) \phi_k(0)=
\sum_{j,k=1}^3\,a_{ij}a_{ik}\,G_{jk}(t),
\eeq
and represents the (approximate) correlation functions of the
eigenstates of the Hamiltonian. 
To check the stability of the procedure we have performed the same
calculation for $t_0=0,\,t=a,2a$ and obtained consistent results in both
cases. The coefficients
$a_{ik}$ take values between zero and one and quantify the overlap
of each individual correlator $G_{ik}$ with the correlators of the
mass eigenstates, $\Gamma_i$. They characterise
the composition of the eigenstates in terms of the original operators
as well as the degree of mixing between operators.

We extract the effective masses 
of the eigenstates $\Phi_i$ by assuming a single exponential 
decay of their correlation functions,
\beq
m^i_{eff}(t)=-\ln\left[\frac{\Gamma_i(r,t+1)}{\Gamma_i(r,t)}\right] \, .
\eeq
The effective masses in lattice
units are so heavy that
the observed ``plateaux'' only extend over two or three 
$t$-values before the signal is lost in noise.
In general we therefore
approximate the $i$-th eigenstate of the static potential in lattice 
units by $aV_i(r/a)=m^i_{eff}(2)$.
This leaves room for a systematic error because 
the asymptotic plateau values might be attained only at larger $t$
and thus the energies be overestimated. 
In the case of the two-gluelump operator this error can be ruled
out because we find it to be fully compatible with twice the 
single gluelump mass, for which we do see a good plateau.
Only for the Wilson loop does this source of uncertainty remain.
Since string breaking occurs when the energies of the broken and unbroken
string are equal, overestimating the latter may lead to 
underestimating the string breaking distance $r_b$.

The size of the operator basis used here is significantly smaller
than in the corresponding three-dimensional study, where up to eight 
operators were used \cite{pw1}. The reason simply is that our present 
computer resouces do not allow for a larger basis in a four-dimensional
calculation. In consequence, whereas we may be fairly confident 
about our mass estimates for the ground state potential, the first and
especially second excited states are likely to be contaminated by 
higher excitations. In particular, since we have only one two-gluelump
operator in our basis, we are unable to compute the first excitation
of the gluelump and thus cannot determine the correct ordering of 
the excited states for distances beyond the breaking scale, $r\gg r_b$.
We shall come back to this point in the discussion of the results.

In order to examine the scaling behaviour of the potential, we have
worked at two values of the bare coupling, $\beta=2.4$ on a
$32^4$ lattice, and $\beta=2.5$ on a $40^4$ lattice.
We did not find a value for $r_0(\beta=2.4)$ in the literature,
and hence based our scaling analysis on the expression
\beq
a(\beta)=400\exp\left[-\frac{\beta\ln 2}{0.205}\right]{\rm fm}\;,
\eeq
which describes the scaling 
of the fundamental representation static 
potential with the lattice spacing in the range of interest \cite{ukqcd}.
For better comparison with other simulations we also quote
our results at $\beta=2.5$ using $r_0(\beta=2.5)=6.39(09)a$ \cite{sommer}.

In three dimensions an explicit test for finite volume effects 
has been performed, with the finding that a spatial size $L\approx 2.4 r_b$
is large enough for $r_b$ to be free from finite size effects \cite{pw1}.
In the present work we shall find that $L > 2.8 r_b$ for both
lattice spacings and we expect no finite size effects.

At $\beta=2.4$ we employed a maximum of 40 smearing steps,
whereas at $\beta=2.5$, 60 steps were necessary to reach approximate 
saturation in the lowering of the effective mass of the Wilson loop.
For both lattice spacings, 50 Monte Carlo sweeps were inserted
between successive measurements, where one heatbath sweep was followed by four
overrelaxed sweeps. For $\beta=2.4$ we collected 400 measurements for the
gauge-invariant approach and 580 for the gauge-fixed off-axis operators,
whereas for $\beta=2.5$ 190 gauge-invariant 
measurements were sufficient to yield comparable
statistical errors. 
The latter were estimated using a jackknife
procedure, except in the case of the off-axis data, where 
we performed a bootstrap analysis to extract errors more reliably.
We have checked the decorrelation of 
our measurements by observing stability of the statistical errors when
the data were binned.

\section{Numerical results}

As was discussed in Section 2, we measured the mass of the gluelump
as a reference to check the projection properties of the two-gluelump
operator. The correlation function $G_G(t)$ is  relatively easy to 
measure as it has a good signal and plateaux in the effective masses
can be identified. For our two lattice spacings we extract
\bea
\label{m_glmp}
aM_G(\beta=2.4)=1.54(1),& aM_G(\beta=2.5)=1.27(1)  \\
M_G(\beta=2.4)=2541(17) {\rm MeV},
& M_G(\beta=2.5)=2938(23) {\rm MeV}
& \left[ r_0 M_G = 8.12(13)\right]. \nn
\eea
In the second line the gluelump mass is given in continuum units. 
Recall however, that there is no physical meaning to these numbers,
but the increase on the finer lattice signals
the linear divergence due to self-interactions of the static source.

\begin{figure}[tbh]
\begin{center}
\leavevmode
\epsfysize=300pt
\epsfbox[20 30 620 730]{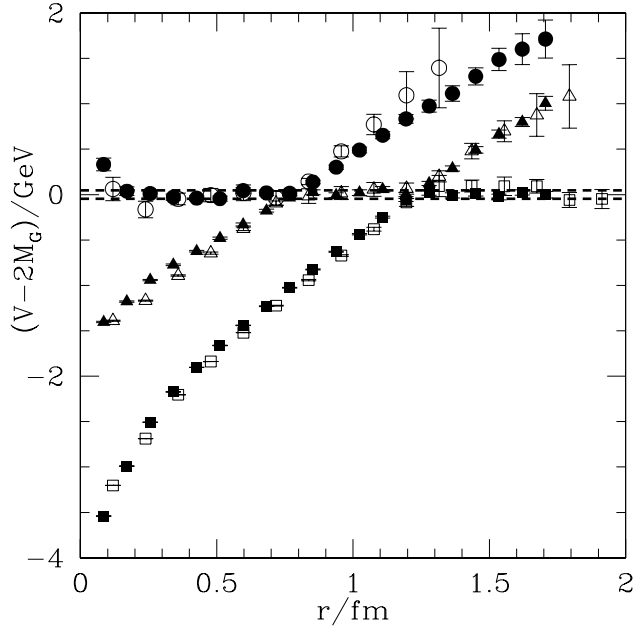}
\end{center}
\vspace{-2.2cm}
\caption[]{\label{pot}
{\it
The adjoint potential and its first two excitations for $\beta=2.4$ (open symbols)
and $\beta=2.5$ (full symbols). The dashed lines mark the error band
for $2M_G$.}}
\end{figure}

\begin{figure}[tbh]
\begin{center}
\leavevmode
\epsfysize=165pt
\epsfbox[20 30 620 730]{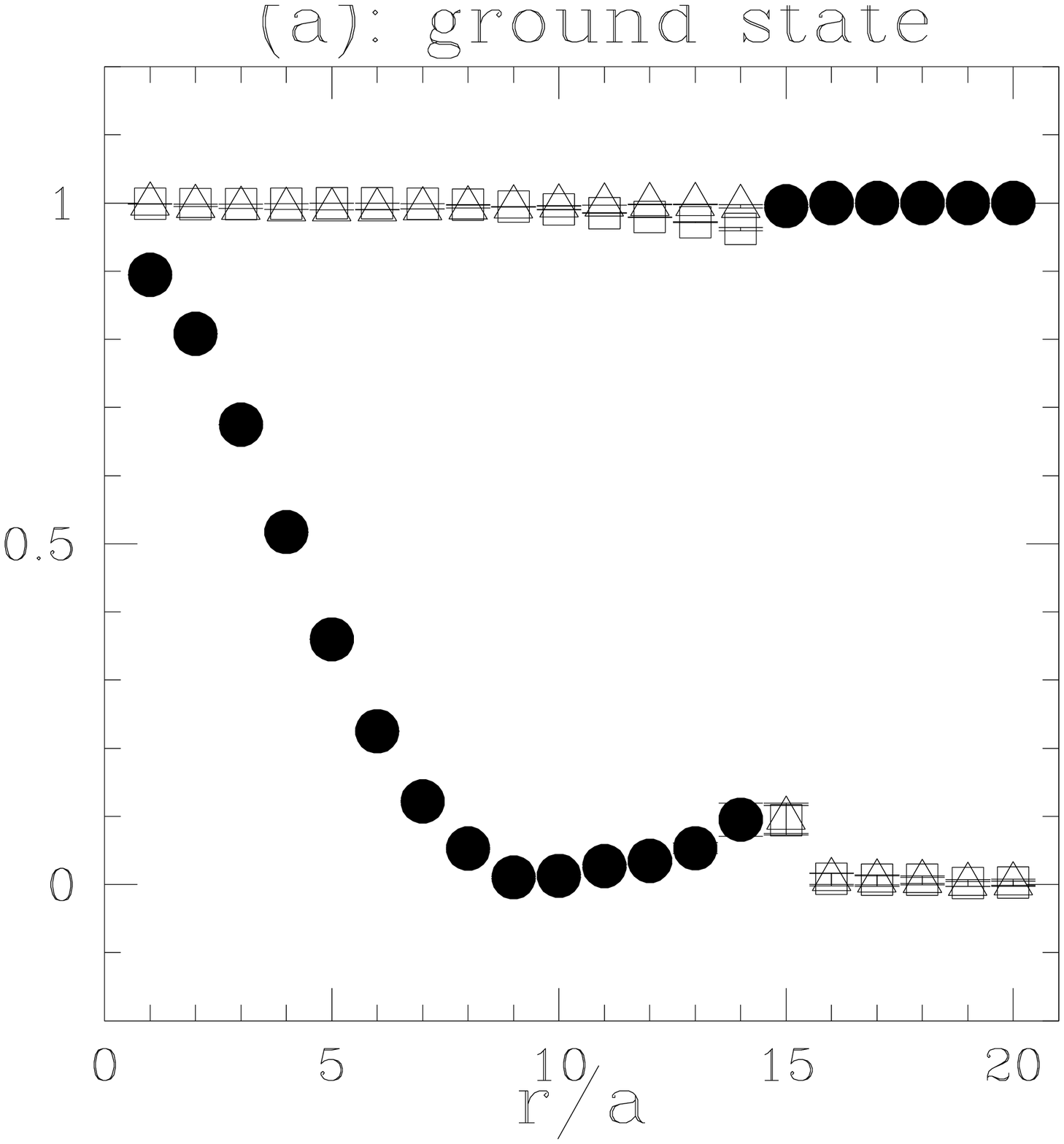}
\leavevmode
\epsfysize=165pt
\epsfbox[20 30 620 730]{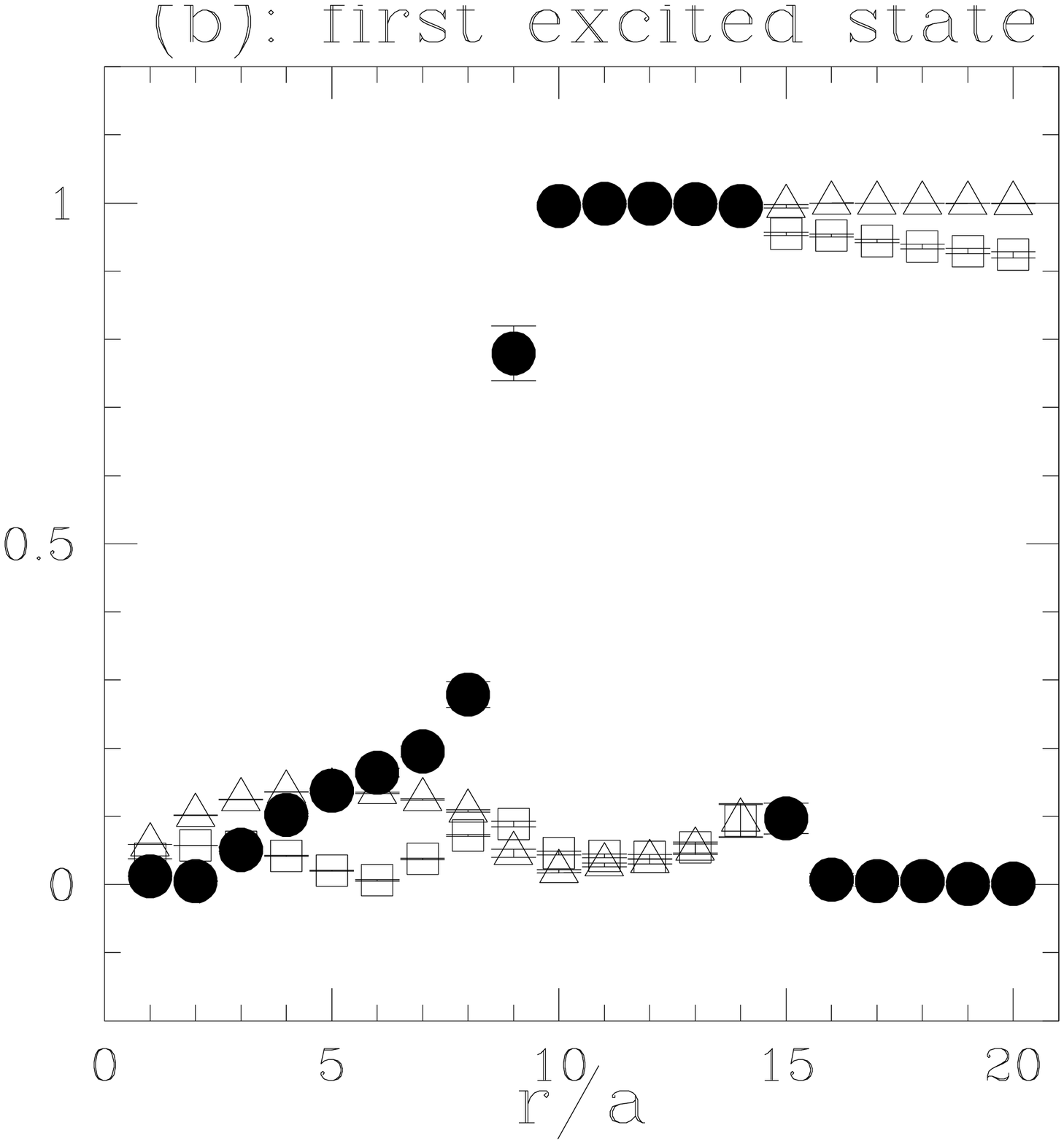}
\leavevmode
\epsfysize=165pt
\epsfbox[20 30 620 730]{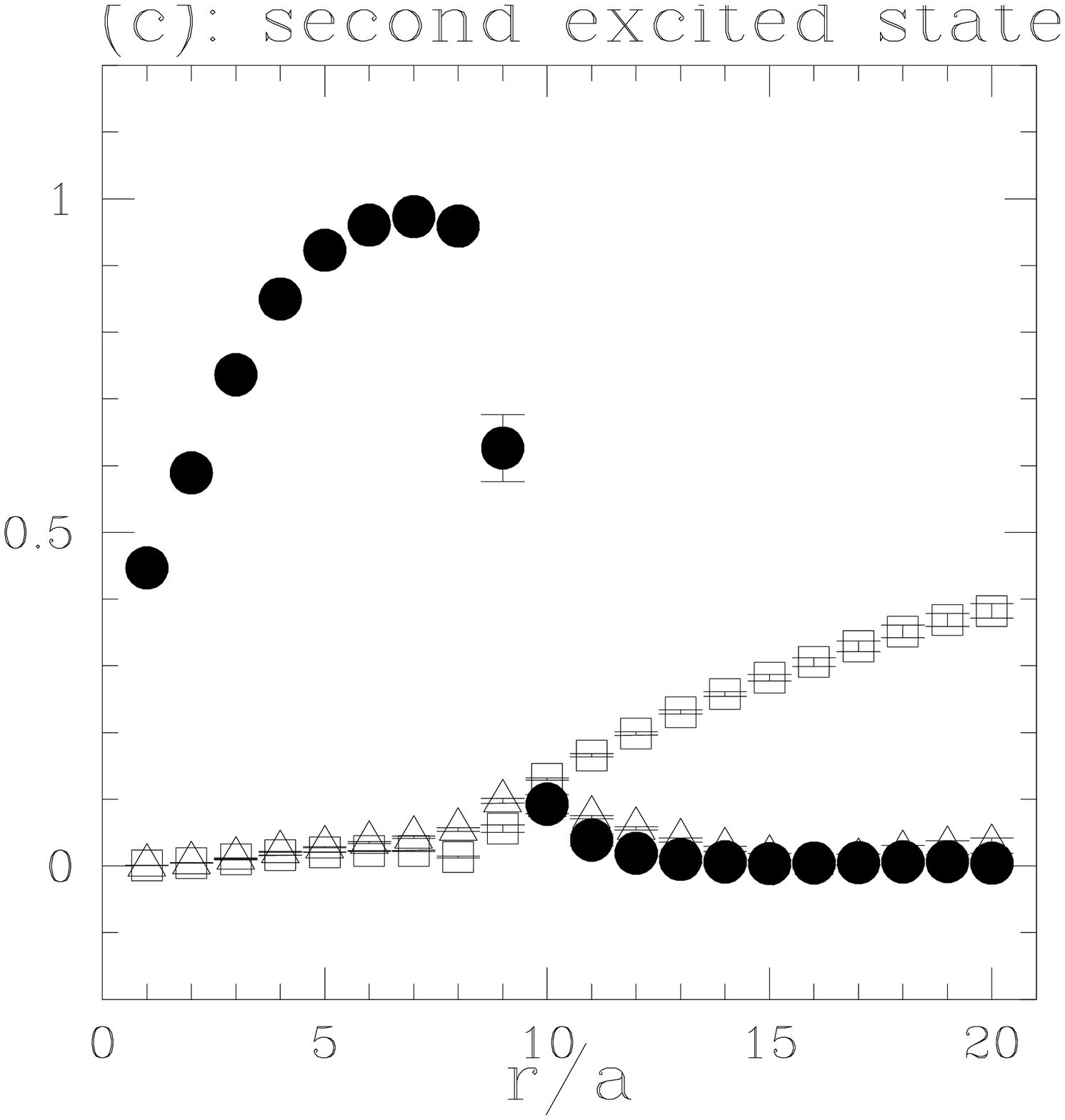}
\end{center}
\vspace{-1.6cm}
\caption[]{\label{ovlp}
{\it The overlaps $a_{ik}$ of the operators in the simulation
with the lowest states of the potential, as obtained 
at $\beta=2.5$. Open triangles and squares: $G_{SS}$ after 40 and 
60 smearings, respectively; Filled circles: $G_{GG}$. 
}}
\end{figure}

Figure \ref{pot} shows the three lowest states of the adjoint static potential
in continuum units, as obtained from the gauge-invariant calculation
at both lattice spacings. We begin by discussing the ground state.
At small distances we observe the expected linear rise corresponding 
to an area law behaviour of the Wilson loop. The ground state then
saturates rather suddenly at $r\approx 1.25{\rm fm}$, 
maintaining an energy value 
of twice the gluelump mass, in accord with the expectation of 
string breaking. This observation is corroborated by the 
analysis of the operator content of the ground state, which is displayed
in Figure \ref{ovlp} (a). In analogy with previous analyses in
(2+1) dimensions \cite{pw1} and the fundamental potential in 
Higgs models \cite{sb1,sb2}, we find the Wilson loop to have nearly
full projection onto the ground state with unbroken string for $r<r_b$,
but practically no projection onto the saturated ground
state for $r>r_b$. 

On the contrary, it maintains its projection onto the unbroken
string state which is still in the spectrum but now as the first excited
state. Conversely, the gluelump operator starts out with some projection onto
the ground state at small distances, which may be due to
gluon exchange between the gluelumps, or simply be an artefact of the
smearing procedure, since smeared gluelump operators overlap for small
distances. However, this overlap with the ground state rapidly decreases
with distance up to the breaking scale, where it projects fully onto
the ground state, the latter now being a two-gluelump state.
The crossover region with appreciable mixing between the two operator types
is even narrower than was reported in the 
three-dimensional calculation, presumably because we are
further from the continuum limit than the simulations in \cite{pw1}.

The ordering of states is completely analogous
to the three-dimensional case \cite{pw1}. 
In particular, at small distances
the first excited state corresponds to an excitation of the gauge string, 
as follows from its linear rise as well as from the overlaps displayed
in Figure \ref{ovlp} (b). It is moreover visible that it receives more 
contributions from the lightly smeared Wilson loop than from the highly 
smeared Wilson loop. Both figures furthermore show quite clearly the
breaking of the excited string. In fact, the observed mixing between
the Wilson loops and two-gluelump operator appears rather more pronounced
than in the very rapid crossover of the ground state string breaking.
We ascribe this to the fact that it happens at smaller $r$, 
where the Wilson loops are not 
as large and thus give
a better signal for the variational calculation.
After the breaking of the excited string, the first excited state of
the potential is the two-gluelump state receiving its main contribution
from the corresponding operator, until this level is attained by the 
ground state string.

For $r>r_b$ we do not see saturation of the linearly rising first excited
state, in contrast to the three-dimensional findings \cite{pw1}.
This is a consequence of our limited basis consisting of just one 
gluelump-pair operator, which prohibits a calculation of 
excited gluelump states. For the same reason, we expect the excited states
to be still contaminated by higher excitations. 

We now follow the prescription given in \cite{pw1} to determine the 
value of the breaking scale $r_b$ in continuum units. A suitable
definition of $r_b$ for this purpose is by the distance where
the string and the two-gluelump state have equal energy,
\beq
  \Delta\equiv E_S-E_{GG}\Big|_{r=r_b} = 0.
\label{eq_delta}
\eeq
To solve this equation we interpolate the energy difference
between the string and the two-gluelump states 
to the point where it vanishes.
Systematic uncertainties are estimated by varying the number of points used
for the interpolation. From this procedure we obtain the results
\bea
r_b/a(\beta=2.4)=10.6(8),& r_b/a(\beta=2.5)=14.4(6) \nn\\
r_b(\beta=2.4)=1.27(8) {\rm fm},& r_b(\beta=2.5)=1.23(5){\rm fm} \,
\left[r_b/r_0=2.25(10)\right]\;,
\label{extra}
\eea
which exhibit satisfactory scaling behaviour.

\begin{figure}[tbh]
\begin{center}
\leavevmode
\epsfysize=300pt
\epsfbox[20 30 620 730]{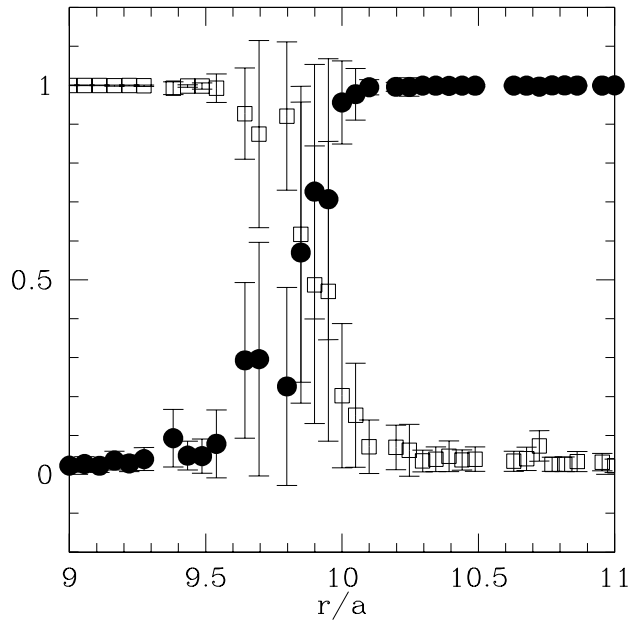}
\end{center}
\vspace{-2.2cm}
\caption[]{\label{offax}
{\it
The overlaps $a_{ik}$ of the operators 
with the ground state of the potential, as obtained
at $\beta=2.4$ in the gauge-fixed calculation. 
Open squares: $G_{SS}$;
Filled circles: $G_{GG}$.
}}
\end{figure}

In order to have a higher resolution in the range of string breaking
and to obtain a better signal
for mixing of the operators in the ground state we have performed
a gauge-fixed calculation 
including off-axis correlations at irrational distances in lattice units.
We have checked explicitly that the on-axis results obtained in this
manner agree with those from a gauge-invariant calculation with 
a $2\times2$ matrix. It turns out, however, that
Wilson loops yield a better signal than the spatial correlation
of gauge-fixed sources, which need more time-like distance $t$
to attain their asymptotic behaviour. The overlaps of the two operators
with the ground state in the string breaking region are shown for 
$\beta=2.4$ in Figure \ref{offax}. Since this is the coarser lattice,
the crossover in the overlaps is very rapid, even with the higher 
resolution. Despite their poorer quality, the data show clear evidence
for progressive mixing between the two operator types, which is absent in 
the gauge-invariant calculation at the same $\beta$, where the ground
state correlation is always composed entirely of either Wilson loop or  
two-gluelump operator. Finally, we remark that the $r_b$ one would
extract from this figure is lower than the one extracted from the
gauge-invariant calculation in eqn.(\ref{extra}), although they are
statistically compatible. We nevertheless interpret this difference
as a systematic effect rather than a statistical one, ascribing it
to the severe limitation that a $2\times 2$ basis poses. 

\section{Conclusions and outlook}

We have measured the adjoint static potential in 4-dimensional $SU(2)$ 
Yang-Mills theory. A scaling study at $\beta=2.4$ and $2.5$ indicates that
string breaking occurs at a distance $r_b \approx 1.25 {\rm fm}, 2.3 r_0$. 
By measuring off-axis correlations in a fixed gauge, we have observed 
mixing between broken and unbroken string operators, over a distance
smaller than one lattice spacing ($\sim 0.1$ fm). We have also resolved
the breaking of the excited string, which occurs at distance 
$\sim 0.75{\rm fm}, 1.4 r_0$.

Our findings are in line with the expectation that the string should break
where the linearly rising part of the potential extracted from pure
Wilson loop calculations intersects with the two-gluelump state
\cite{mic92}. The suddenness of the breaking, characterized by the
near-absence of mixing noted above, as well as the ordering of the 
excitations are similar to what happens in 
$(2+1)$ dimensions \cite{pw1}. In fact, the similarity is
even quantitative: the string-breaking distance in units of the lightest
glueball mass \cite{teper4d} is practically the same ($r_b m_{0^{++}} \sim 9.7$ 
versus $10.3(\pm1.5)$ in \cite{pw1}). This suggests that identical effective 
mechanisms are at work.

We find no reason to expect surprises in the case of $SU(3)$. At $\beta=6.0$,
the gluelump mass is known ($aM_G=1.33$ \cite{UKmichael}), and string breaking
should occur at $r_b \sim 12.5 a$ \cite{UKmichael}, which corresponds again
to a similar value of $r_b m_{0^{++}}$. To observe adjoint string breaking
at $\beta=6.0$, a $32^4$ lattice should be marginally sufficient.
Such a project is well within reach of better-equipped lattice groups.

The only technical difficulty here is to ascertain that broken and unbroken
string operators are not completely decoupled, but do mix in the close
vicinity of $r_b$. The mixing ``window'' is remarkably narrow, 
and it would be interesting to find the dynamical reasons for this. 
According to a recent suggestion \cite{provero}, the
suddenness of string-breaking might be caused by a qualitative change
in the world-sheet spanned by the 2 static charges, which tears abruptly
as the two sides are taken apart. If this change is of topological nature, 
one may wonder whether mixing should occur at all, or whether the string
simply ``snaps''.

\paragraph{Acknowledgements}

The calculations for this work were performed on the NEC-SX4/32 at the
HLRS Universit\"at Stuttgart and on the CRAY-SV1 at ETH Z\"urich.
O.P. also wishes to thank the ITP Universit\"at Heidelberg for continued
access to their workstations, where much of the current analysis was done.

\end{document}